\documentclass[anonymous=false, %
               format=acmsmall, %
               review=false, %
               screen=true, %
               nonacm=true]{acmart}

\usepackage[ruled]{algorithm2e}
\usepackage{amsmath}
\usepackage{color}
\usepackage{hyperref}

\usepackage{listings}
\newcommand{\eps}{\varepsilon}
\newcommand{\RR}{\mathbb{R}}

\newcommand{\AR}{\text{AR1}}

\usepackage{parskip}

\begin{document}

\title{Structural time series grammar over variable blocks}

\author{David Rushing Dewhurst}
\affiliation{%
  \institution{Charles River Analytics}
  \streetaddress{625 Mount Auburn St}
  \city{Cambridge}
  \state{MA}
  \postcode{02138}
  \country{USA}
}
\email{ddewhurst@cra.com}

\begin{abstract}
A structural time series model additively decomposes into generative, semantically-meaningful 
components, each of which depends on a vector of parameters.
We demonstrate that considering each generative component together with its 
vector of parameters as a single latent structural time series node can simplify reasoning about collections of structural time series components. 
We then introduce a formal grammar over structural time series nodes and parameter vectors.
Valid sentences in the grammar can be interpreted as generative structural time series models.
An extension of the grammar can also express structural time series models that include changepoints, though these models are necessarily not generative. 
We demonstrate a preliminary implementation of the language generated by this 
grammar.
We close with a discussion of possible future work.
\end{abstract}

\maketitle

\section{Introduction}

Structural time series (STS) are interpretable time series models that are widely used in 
economics \cite{choi2012predicting},
 finance \cite{dossche2005measuring},
 marketing \cite{brodersen2015inferring},
  and
  climate science \cite{rodionov2006use}.
These models posit an additive decomposition of observed time series into 
multiple latent time series, each of which often has a simple interpretation.
Each latent time series
depends on a vector of parameters, optimal values of which are learned during 
inference.
STS model expressiveness can be increased by adding more latent components or by 
replacing static parameter vectors with vectors of time-dependent latent components.

Here, we explore a method for reasoning about STS models by grouping each latent time series together with its vector of parameters into a single variable block.
This method corresponds to a choice of joint density factorization and can simplify graphical displays of STS models.
We then define a grammar that formalizes the process of addition and function composition of STS variable blocks. 
An extension to the grammar allows for expression of changepoint models. 
We outline a preliminary implementation of the
resulting language using a small, diverse set of variable blocks, and
demonstrate the variety of dynamic behavior generated by models corresponding to 
strings in the grammar.
We close with suggestions for future work.
We propose the formalization of the grammars introduced here and the creation of a domain-specific structural time series language (STSL), along with extensions to multiple observed time series.

\section{Structural time series}\label{sec:sts}
We define a structural time series (STS) as a time series model of the form
\cite{moore_structural_nodate,choi2012predicting}
\begin{equation}\label{eq:sts-basic}
y(t) = \eps(t; \sigma) + \sum_{k=1}^K f_k(t; \theta_k),
\end{equation}
where $\theta_k \in \RR^{N_k}$ is the vector of parameters of the function $f_k$. 
The noise term $\eps(t;\sigma)$ is a draw from a zero-mean location-scale-family probability distribution $p(\eps|0, \sigma)$.
Usually $p(\eps|0, \sigma)$ is taken to be a normal distribution.
We will follow this convention because it has the convenient property that $\sum_n \eps_n(t; \sigma)$ is again normally distributed.
The only observed random variable is $y(t)$; all other random variables (rvs) are latent.

We now outline two illustrative examples of STS models and describe how they can be
modified through the operations of addition and function composition.

STS models can be extended through addition of multiple components. 
A simple time series model is a linear regression in time, $f(t) = a_0 + a_1 t +
 \eps(t; \sigma)$, known as a global trend model.
A model similar to Facebook's ``Prophet'' \cite{taylor2018forecasting} extends this simple model by adding seasonality and irregularity terms to the global trend,
$$
f(t) = \eps(t) + a_0 + a_1 t
+ 
	\sum_{s=1}^S \gamma_{s}\mathbb{I}_{s \text{ mod }t\ =\ 0}(t)
+ v(t),
$$
where $v(t)$ is defined by an order-1 autoregression, $v(t) = \beta v(t - 1) + \xi(t; \nu)$
with $\xi(t;\nu ) \sim \text{Normal}(0, \nu^2)$.
More terms could be added to this model to capture other temporal phenomena, e.g., a term $h(t)$ that specifically captures holiday effects \cite{taylor2018forecasting}.

STS models can also be extended through function composition. 
The simplest STS model is pure white noise, 
$y(t) = \eps(t; \sigma)$,
where $\eps(t; \sigma) \sim \text{Normal}(0, \sigma^2)$.
In financial applications $y(t)$ can represent the  instantaneous ``return'' on an asset,
or log difference (roughly equivalent to percent change) in asset price
\cite{black1973pricing}.
It is established that this naive model does not accurately describe observed asset return dynamics \cite{gatheral2006volatility}.
A modified version of this model allows the standard deviation of $y(t)$, also called volatility, 
to change in time:
\begin{subequations}
\begin{align}
\log \sigma(t) &= \log \sigma(t - 1) + \xi(t; \nu)  \label{eq:stoch-vol-1}\\
y(t) &= \eps(t; \sigma(t)) \label{eq:stoch-vol-2},
\end{align}
\end{subequations}
where $\xi(t; \nu) \sim \text{Normal}(0, \nu^2)$ and $\nu > 0$.
Other parameters such as $\nu$ could also be replaced with time dependent components
to increase model expressiveness.

\section{Block structure}
\begin{figure}
	\centering
	\includegraphics[width=0.75\textwidth]{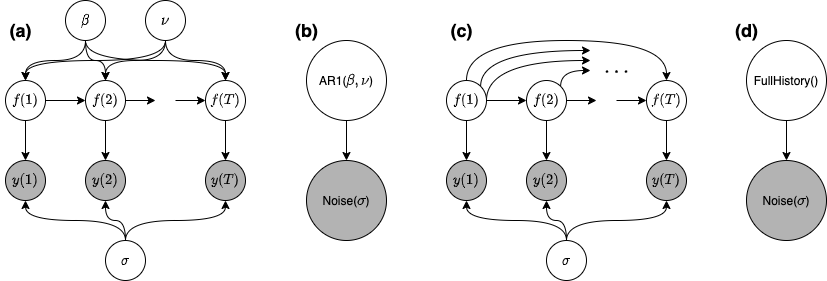}
	\caption{
		An STS block can simplify the graphical representation of model structure.
		Panel (a) displays the graphical representation of an autoregressive process,
		while panel (b) displays an alternate ``block-based'' representation of the same process. In the block-based representation, the recursive structure of the latent AR1 process has been rolled up into a single latent node. 
		In panel (c) we display a graphical representation of a full-history latent process. 
		This process is non-Markov and difficult to represent using typical graphical notation. 
		However, if there exists an explicit algorithm to compute the latent process, 
		it can again be represented in a single latent node, as displayed in panel (d).
	}
	\label{fig:plate-vs-block}
\end{figure}

It is useful to group the parameters and variables of an STS component into a semantically-meaningful ``block'' 
that can be reasoned about as a single entity.
Causal relationships between STS components can be reasoned about more clearly using these groupings and the induced block structure can provide a cleaner graphical display, e.g.,
in a plate diagram.
We now introduce the block structure with a motivating example. 
Consider the STS model defined by a basic autoregressive state-space process:
\begin{subequations}
	\begin{align}
f(t) &= \beta f(t - 1) + \xi(t; \nu) \label{eq:block:ar}\\
y(t) &= \eps(0; \sigma) + f(t) \label{eq:block:ar-obs}
	\end{align}
\end{subequations}
A usual plate diagram would represent Eq.\ \ref{eq:block:ar} as an unrolled set of nodes
each causally influencing the value of the next as demonstrated in panel (a) of Fig.\ \ref{fig:plate-vs-block}.
However, we could instead group the recursive definition of $f(t)$,
 along with the parameters $\beta$ and $\nu$, into a single object called $\AR(\beta, \nu)$.
This object, which we refer to as an AR1 block, describes both the process that generates the latent time series $f(t)$ and the vector of parameters 
$\theta = (\beta, \nu)$ that are used in the data generating process.
We demonstrate this grouping in panels (a) and (b) of Fig.\ \ref{fig:plate-vs-block}.

The block notation corresponds to an explicit way of factoring of the joint density function. Let $x(t)$be the single scalar observed time series,
 $z(t) = (z_1(t), z_2(t), ..., z_K(t))$ the vector of all latent time series at time $t$, and
 $u$ the vector of all global rvs.
The usual graphical representation of a model, as in Fig.\ \ref{fig:plate-vs-block} panels
(a) and (c), corresponds to a factoring of the joint density as
\begin{equation*}
p(x, z, u) = p(u) \prod_t p(x(t)|x( t' < t), z(t' \leq t), u)\prod_i p(z_i(t)|z_{\lnot i}(t' < t), u).
\end{equation*}
By $z(t' < t)$ we mean the time series of rvs $z(1), z(2), ..., z(t' - 1)$. 
The notation $z_{\lnot i}(t)$ represents the vector $(z_1(t),..., z_{i-1}(t), z_{i+1}(t),...)$.
In this factorization, each random variable at each timestep is explicitly represented by a likelihood or prior term and corresponds to a unique node in the graphical representation. 
The block notation instead corresponds to the factorization
\begin{equation}
p(x, z, u) = p(x|z, u) p(u)  \prod_i  p(z_i |z_{\lnot i}).
\end{equation}
The temporal relationships between $z_i(t)$ and $z_i(t')$ for $t' < t$ are now implicit and defined
within the prior terms $p(z_i|z_{\lnot i})$. 
Similarly the dynamic structure of the likelihood function is now implicit in $p(x|z, u)$ and 
is absent from the graphical representation. 
The choice to use block notation thus shifts model complexity from edge space to node space: the relationship between a pair of time series is represented by a single edge in the 
graph, but the definition of the multivariate random variable defining each node is 
more complicated.

\section{STS grammar}
We introduce a grammar over STS models,  $G = (V, \Sigma, R, S)$,
where 
$V = \{S, Q, p\}$, $\Sigma = \{f, \theta\}$, and the production rules $R$ are
\begin{subequations}
\begin{align}
S &\rightarrow Q\ |\ Q + S \label{eq:grammar:addition} \\
Q &\rightarrow f(p)  \label{eq:grammar:block-like}\\
p &\rightarrow \theta\ |\ (S\ |\ p,...,S\ |\ p) \label{eq:grammar:parameters}
\end{align}
\end{subequations}
The nonterminal symbols $S$, $Q$, and $p$ represent an STS model, a block object, 
and a parameter-like object respectively. 
The symbol $Q$ is redundant
but we include it because it makes a later expansion of $G$ to include changepoints easier.
The terminal symbols $f$ and $\theta$ represent the generative component of a block and 
the parameters of the generative component of a block respectively. 
Production rules \ref{eq:grammar:addition} and \ref{eq:grammar:block-like} jointly state that any STS can be extended by adding another block.
Production rule \ref{eq:grammar:parameters} states that $p$ can be replaced with either an $N$-dimensional parameter vector $\theta$ or by any combination of STSs and parameter-like objects that satisfy the constraints imposed by the associated block.
The dimensionality $N$ is equal to the dimensionality of the parameter space of the 
associated generative block component.

We can extend $G$ to express changepoint models using an
augmented grammar $G' = (V, \Sigma, R', S)$.
We define $R'$ by replacing the second rule of $R$ with 
\begin{equation}
Q \rightarrow f(p)\ |\ C(S, f(p))\ |\ C(f(p), S).
\end{equation}
The symbol $C$ is the changepoint operation; $C(x, y)$  yields a new block defined by the 
concatenation of $(x(1), ..., x(t^* - 1))$ and $(y(t^*),...,y(T))$ at a random time point {}$t^* \sim 
\text{DiscreteUniform}(\{2,...,T~-~1\})$.
Models corresponding to sentences in $G'$ are not generative because the existence of changepoints means that an end time must be imposed and all values of the observed 
time series must be known between the start and end times. 

This grammar could be used to define a domain specific language (DSL).
This DSL would be an interface to easily describe STS models and express relationships between latent and observed time series (for a discussion of extending the definition of STS models to include multiple observed
time series, see Sec.\ \ref{sec:future-work}).

\section{Implementation}

We have implemented the basic functionality of the language generated by grammar $G'$ in the host PPL Pyro \cite{bingham2019pyro}.
Our implementation is available online\footnote{
	\href{https://gitlab.com/daviddewhurst/stsb}{https://gitlab.com/daviddewhurst/stsb}
}. There is a large range of dynamic behaviors expressible through the 
small number of operations and STS blocks that we have implemented,
though our implementation is preliminary.

We have implemented a small set of fundamental blocks (corresponding to $f$ in 
grammars $G$ and $G'$) that capture a range of potential time series phenomena:
random walk and geometric random walk, $f(t) = f(t - 1) + \eps(t; \sigma)$ and $f(t) = \exp g(t)$ with $g(t)$ a random walk;
first order autoregression, $f(t) = \beta f(t - 1) + \eps(t; \sigma)$;
seasonality, $f(t) = \cos (2 \pi t / \rho)$;
global trend, $f(t) = a_0 + a_1 t$;
a zero model, $f(t) = 0$;
and a non-Markov block, $f(t) = F[f(t - 1), f(t - 2),...,f(1); \eps(t, \sigma)]$.
The function $F$ in the non-Markov block is an arbitrary user-defined function 
of all past values of the latent time series $f(t)$, the current time value $t$, another 
time argument $s < t$ (which could be used for, e.g., constructing convolution 
operations); and a noise rv $\eps(t; \sigma)$.
\begin{figure}
	\centering
	\includegraphics[width=0.8\textwidth]{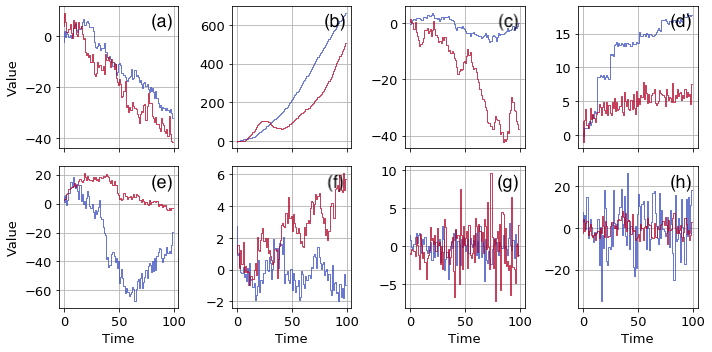}
	\caption{
		We display draws from prior predictive distributions for various STS models expressible 
		using grammar $G'$. 
		The models are: random walk (panel a);
		 local linear trend (panel b);
		semi-local linear trend (SLLT) (panel c);
		full-history non-Markov model (panel d);
		changepoint model with SLLT generative components (panel e);
		seasonal-global trend (panel f);
		stochastic volatility (panel g);
		and additive seasonality model with changepoint (panel h).
		These models can be generated using the library at 
		\href{https://gitlab.com/daviddewhurst/stsb}{https://gitlab.com/daviddewhurst/stsb}.
	}
	\label{fig:sts-examples}
\end{figure}
\begin{table}
	\centering
	\begin{tabular}{| l | l | } \hline
		\textbf{Description} & \textbf{Expression} \\ \hline
		Random walk & 
		\begin{lstlisting}[boxpos=t]
rw = Noise(
	RandomWalk(t1=t1),
	t1=t1
)
		\end{lstlisting} \\ \hline
	Local linear trend & 
	\begin{lstlisting}[boxpos=t]
llt = Noise(
	RandomWalk(
		loc=RandomWalk(t1=t1),
		t1=t1,
	),
	t1=t1
)
	\end{lstlisting} \\ \hline
	Changepoint model & 
	\begin{lstlisting}[boxpos=t]
sllt1 = RandomWalk(loc=AR1(t1=t1), t1=t1)
sllt2 = RandomWalk(loc=AR1(t1=t1), t1=t1)
changepoint_sllt = Noise(
	changepoint_op(sllt1, sllt2), 
	t1=t1
)
	\end{lstlisting} \\ \hline
	Stochastic volatility & 
	\begin{lstlisting}[boxpos=t]
stoch_vol = Noise(
	Zero(t1=t1),
	t1=t1,
	scale=GeometricRandomWalk(t1=t1)
)
	\end{lstlisting} \\ \hline 
	Multiple seasonality \\
	with changepoint & 
	\begin{lstlisting}[boxpos=t]
s = [Seasonal(t1=t1) for _ in range(4)]
random_seasonal = Noise(
	s[0] + s[1] + changepoint_op(s[2], s[3]),
	t1=t1
)
	\end{lstlisting} \\ \hline
	Stochastic optimization\\
	null model & 
	\begin{lstlisting}[boxpos=t]
fn = lambda t, s, y, noise: torch.cat((
	y.max().view((1,)), 
	y.median() + noise.view((1,))
)).max()
model_improvement = NonMarkov(t1=t1, fn=fn)
optim_null_model = Noise(model_improvement, t1=t1)
	\end{lstlisting} \\ \hline
	\end{tabular}
\caption{
	We display example implementations of various STS models described using grammar $G'$.
	The current host language is Python and host PPL is Pyro. 
	It is possible to express multi-part or full-history models, such as the multiple seasonality model or stochastic optimization null model, in very few lines of code.
	A minimal implementation of the language generated by $G'$ is at 
	\href{https://gitlab.com/daviddewhurst/stsb}{https://gitlab.com/daviddewhurst/stsb}.
}
\label{tab:examples}
\end{table}

In Table \ref{tab:examples} we display sample Python code implementing multiple STS models expressed using the syntax of $G'$:
 a latent random walk model (panel a);
 a local linear trend, which is a random walk with mean given by another random walk process (panel b);
 a nonstationary changepoint model (panel e);
 the stochastic volatility model represented by Eqs.\ \ref{eq:stoch-vol-1} and
 \ref{eq:stoch-vol-2} (panel g);
 a model incorporating multiple added seasonal components and a changepoint
 (panel h);
 and a full-history non-Markov model that could be used as a null model of
 objective function value during stochastic optimization (panel d).
 In Fig.\ \ref{fig:sts-examples} we display sample draws from the prior predictive distributions of each of these STS models and also display sample draws from two additional models: a semi-local linear trend, which is a random walk with mean given 
 by an AR1 process (panel c); and a seasonal-global trend model similar to that
 described in Sec.\ \ref{sec:sts} (panel f).

\section{Future work}\label{sec:future-work}

Future work should proceed in multiple directions.
\begin{itemize}
	\item[1.] The grammars $G$ and $G'$ should be formalized and a DSL should be created
	to implement the language generated by $G'$.
	\item[2.] Ideally, the DSL would support more blocks than we have currently implemented. Useful blocks include: discrete seasonality; other non-Markov processes (e.g., self-exciting point processes, pantograph processes); discrete-valued processes;
	and switching processes.
	\item[3.] The definition of an STS to include only one observed time series and the restriction of the observed and latent time series to be one-dimensional are unnecessarily restrictive. The implementation of the DSL should allow for multiple observed time series and multivariate time series.
	\item[4.]  We envision the DSL being used in the task of 
	latent time series structure search:
	 finding the DAG defined on STS block nodes 
	 that best describes the observed time series. 
	This task will be difficult because of the large, growing search space.
	It may be possible to map this task onto a combinatorial multi-armed bandit problem
	and find an approximately optimal solution 
	\cite{chen2013combinatorial}.
\end{itemize}

\begin{acks}
We are grateful for useful conversations with Avi Pfeffer.
\end{acks}

\bibliographystyle{acm-reference-format}
\bibliography{probprog-2020}

\end{document}